\pdfoutput=1

\documentclass[twocolumn]{aastex631}

\usepackage{amsmath}
\usepackage{calligra}
\usepackage{units}
\usepackage[normalem]{ulem} 

\DeclareMathAlphabet{\mathcalligra}{T1}{calligra}{m}{n}
\DeclareFontShape{T1}{calligra}{m}{n}{<->s*[2.2]callig15}{}
\newcommand{\scriptr}{\mathcalligra{r}\,}

\begin{document}

\title{Strong Bow Shocks: Turbulence and An Exact Self-Similar Asymptotic}

\correspondingauthor{Marcus DuPont}
\email{md4469@nyu.edu}

\makeatletter
\patchcmd{\ltx@foottext}{%
  .5\textwidth\advance\hsize-18pt}{%
  \linewidth\advance\hsize-1.8em%
}{}{}
\makeatother

\author[0000-0003-3356-880X]{Marcus DuPont}
\affiliation{Center for Cosmology and Particle Physics\\
New York University, New York, NY, 10003, USA}

\author{Andrei Gruzinov}
\affiliation{Center for Cosmology and Particle Physics\\
New York University, New York, NY, 10003, USA}

\author[0000-0002-0106-9013]{Andrew MacFadyen}
\affiliation{Center for Cosmology and Particle Physics\\
New York University, New York, NY, 10003, USA}



\defcitealias{Wilkin+1996}{W96}
\defcitealias{Yalinewich+2016}{YS16}
\newcommand\W{\citetalias{Wilkin+1996}}
\newcommand\YS{\citetalias{Yalinewich+2016}}
\newcommand{\A}{\mathcal{A}}
\newcommand{\V}{\mathcal{V}}
\newcommand{\C}{\mathcal{C}}
\newcommand{\B}{\mathcal{B}}
\def \be {\begin{equation}}
\def \ee {\end{equation}}
\newcommand\ag[1]{{\textcolor{red} {\bf #1}}} 
\newcommand\am[1]{{\textcolor{blue} {\bf #1}}} 
\begin{abstract}
We show that  strong 
bow shocks are turbulent and non-universal near the head, but asymptote to a universal steady, self-similar, and analytically solvable flow in the downstream. The turbulence is essentially 3D, and has been confirmed by a 3D simulation. The asymptotic behavior is confirmed with high resolution 2D and 3D simulations of a cold uniform wind encountering both a solid spherical obstacle and stellar wind.  This solution is relevant in the context of: (i) probing the kinematic properties of observed high-velocity compact bodies --- e.g., runaway stars and/or supernova ejecta blobs --- flying through the interstellar medium; and (ii) constraining stellar bow shock luminosities invoked by some quasi-periodic eruption (QPE) models.
\end{abstract}



\section{Introduction} \label{sec:intro}
Astrophysical bow shocks are ubiquitous; they form
at stars flying supersonically through the interstellar medium \citep[e.g.,][]{vanBuren+1988,Kobulnicky+2010,Mackey+2012,Mackey+2015}, around planets \citep[see e.g.,][for comprehensive reviews]{Tsurutani+1985,Truemann+2008}, or colliding wind binaries \citep[e.g.,][]{Stevens+1992,Myasnikov+1998,Gayley+2009,Sasaki+2024}, to name a few scenarios. 

Astrophysical bow shocks can be magnetized and collisionless, but we will study them in the hydrodynamics approximation. Conserved mass, momentum, and energy make the  hydrodynamics approximation meaningful even outside of its precise domain of applicability. 

\citet[][hereafter \W]{Wilkin+1996} gives an exact analytic solution for a stellar wind bow shock ``without pressure'', corresponding to infinitely efficient cooling. Far from the bow shock head, \W{}'s bow shock asymptotes to $z \propto r^3$. The \W{} solution was later generalized to wind-wind shocks including the case where one of the winds is anisotropic \citep{Tarango-Yong+2018}. 

The bow shock  with infinite Mach number is a classic problem dating back many decades with analytical formulations discussed in the works of \citet{Chester+1956,Chester+1956b}, \cite{Freeman+1956},\cite{vanDyke+1957}, \cite{Hayes+Probstein+1959}, and \cite{Yakura+1962}, to name a few of the earliest works in the hydrodynamics approximation.\@ However, even magnetospheric bow shocks can be studied in the hydrodynamic limit given that the Alfv\'en and slow rotational magnetosonic waves remain small while the magnetohydrodynamic Mach number is large \citep{Spreiter+1966,Spreiter+Stahara+1980,Slavin+Holzer+1981,Verigin+2003,Kotova+2020,Kotova+2021}.
In this work, we are interested in the asymptotic shape of the bow shock far from the obstacle in the astrophysical context, and we highlight the steady-state solution that was explored with direct hydrodynamical simulation in \citet[][hereafter \YS]{Yalinewich+2016}. To find the shape of the bow shock, \YS{} propose the thought experiment:  replace an object moving through a cold medium  by a properly timed series of explosions. Assuming that the resulting blast waves are sandwiched between each other,  \YS{} get the bow shock shape $z \propto r^2$.  \YS{} was an important motivation for us, although formally speaking we do not use any of their results. Unlike \YS{}, we derive an \emph{exact} self-similar asymptotic for the strong bow shock, and we also discuss turbulence that necessarily manifests near the obstacle head.

Our main interest is the astrophysically relevant case of a bow shock created by the stellar wind, although we also consider a rigid obstacle. In this work, we prove that the non-relativistic non-radiative bow shock asymptotes to $z \propto r^2$ and give the exact expressions for the flow in the far asymptotic. We confirm these exact expressions by high-resolution hydrodynamical simulations.  We also prove that a bow shock from the stellar wind is turbulent near the head. The turbulence is essentially 3D. We confirm the turbulence by 3D hydrodynamical simulations.

Theoretical considerations are in Section \ref{sec:formalism}. Numerical results are in Section \ref{sec:results}. We conclude and discuss the relevance of our work in Section \ref{sec:discussion}.

\section{Theory}\label{sec:formalism}
One would think that a mathematical description of a bow shock created by a stellar wind is insurmountably difficult. There are at least three surfaces of discontinuity in the flow: two shocks and one tangential discontinuity\footnote{The terminology and notations of this section follow \cite{Landau+Lifshitz+1959}.},
see Figure \ref{fig:sketch}. The shapes of these singular surfaces are determined by the flow in between them, and the flow itself depends on the shapes of the singular surfaces. ``Mathematically such problems are altogether inaccessible to our present analytical techniques'' wrote von Neumann before presenting his exact solution of the strong explosion problem \citep{vonNuemann+1947}.

Moreover, we will {\it prove} that the bow shock is turbulent, and surely one doesn't have the mathematical description of turbulence, yet --- in the far downstream --- the flow admits an exact solution for the same reason that the strong explosion problem admits an exact solution at large time after the explosive material has transferred most of its energy to the atmospheric gas. 

 In Section~\ref{sec:turb} we prove that the bow shock is turbulent. In Section~\ref{sec:analytics} we derive the exact far-downstream asymptotic. The theoretical considerations presented below are very simple -- hard to believe they do work, given that ``such problems are altogether inaccessible'', but Section~\ref{sec:results} fully confirms this simple theory by direct high-resolution axisymmetric 2D and full 3D numerical simulations. 

\subsection{Turbulence}\label{sec:turb}

The bow shock must be turbulent near the head, at least if a bow shock is created by the stellar wind. Here is the proof.

\begin{figure}
    \centering
    \includegraphics[width=0.9\columnwidth]{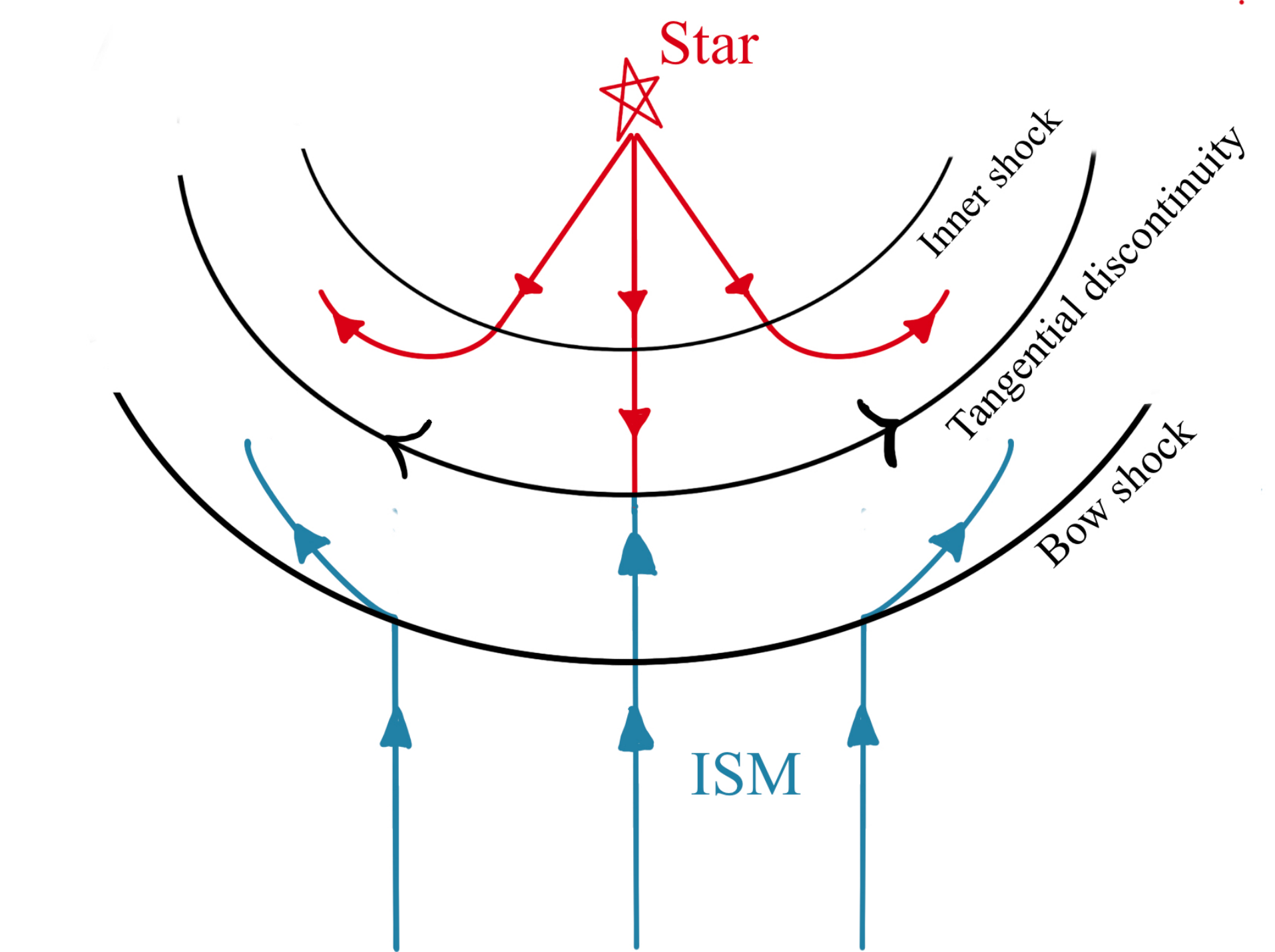}
    \caption{Pictorial depiction of a stellar wind colliding with the cold interstellar medium, which forms a bow shock and inner shock with a tangential discontinuity separating the two shocks.}
    \label{fig:sketch}
\end{figure}

Assume that the flow is laminar. Then, in the frame of the star, the flow is steady.
In this frame, the cold stellar wind collides with the cold interstellar medium
(ISM) wind, see Figure \ref{fig:sketch}. Before colliding, the two winds shock and pressurize. The pressurized ISM and the pressurized stellar gas touch each other along a steady tangential discontinuity surface. The gas velocity normal to the tangential discontinuity is zero, the pressure doesn't jump across the discontinuity, but the tangential velocity does jump. In {\it three} space dimensions, even a finite-Mach tangential discontinuity is unstable. According to \citep[][section 84, problem 1]{Landau+Lifshitz+1959}, this was shown by \cite{Syrovatskii+1954}. In an ideal inviscid gas, the instability of the tangential discontinuity manifests as turbulence. The instability and the resulting turbulence are essentially 3D; we will need a 3D simulation to confirm turbulence in Section~\ref{sec:results}.

\subsection{Asymptotically Self-Similar Bow Shocks}\label{sec:analytics}

We have just proved that the flow must be turbulent near the head of the bow shock. Nevertheless we will now assume that turbulence dies out in the far downstream, and the flow becomes steady. 

Equations of steady gas flow -- continuity, Euler, and adiabaticity -- for the density $\rho$, velocity ${\bf v}$, and sound speed $c$ ($c^2=\gamma \frac{p}{\rho}$, $p$ is the pressure) are
\begin{gather}
    \nabla \cdot(\rho{\bf v})=0\label{gf1},\\
    \gamma \rho({\bf v}\cdot \nabla){\bf v}+\nabla(\rho c^2)=0,\label{gf2}\\
    ({\bf v}\cdot \nabla)\left(\ln c-\frac{\gamma -1}{2}\ln \rho\right)=0,\label{gf3}
\end{gather}
where $\gamma$ is the adiabatic index, assumed to be constant.

Assume an axisymmetric flow without a swirl: in cylindrical coordinates $(r,\phi,z)$, the velocity vector is ${\bf v}=(u,0,v)$ and $\rho$, $c$, $u$, $v$ are functions of $r$ and $z$ only. The shock is at $r=R(z)$. The incoming flow is $\rho=1$, $c=0$, $u=0$, and $v=1$. 

Equations~\eqref{gf1}~--~\eqref{gf3} should then be solved in the domain $r<R(z)$ with the boundary conditions at $r=R(z)$ which follow from the shock jump conditions:
\begin{gather}
    \rho=\frac{\gamma +1}{\gamma -1},\label{bc1}\\
    c=\frac{\sqrt{2\gamma(\gamma -1)}}{\gamma +1}\frac{R'}{\sqrt{1+R'^2}},\label{bc2}\\
    u=\frac{2}{\gamma +1}\frac{R'}{1+R'^2},\label{bc3}\\
    v=\frac{1+\frac{\gamma -1}{\gamma +1}R'^2}{1+R'^2}\label{bc4}.
\end{gather}

As written, Equations~\eqref{gf1}~--~\eqref{gf3} with the boundary conditions \eqref{bc1}~--~\eqref{bc4} have no meaningful solutions (except $R'=\infty$ --- a strong planar shock), because the obstacle creating the bow shock is not specified. We will show, however, that at $z\rightarrow +\infty$ a self-similar non-singular {\it asymptotic} solution does exist. Namely, for $z\rightarrow +\infty$, one can use the following expansion in small $R'$:
\be\label{as1}
\rho=\rho_0(\xi)+\rho_2(\xi)R'^2+\rho_4(\xi)R'^4+...,
\ee
\be\label{as2}
c=c_1(\xi)R'+c_3(\xi)R'^3+c_5(\xi)R'^5+...,
\ee
\be
u=u_1(\xi)R'+u_3(\xi)R'^3+u_5(\xi)R'^5+...,
\ee
\be\label{as4}
v=v_0(\xi)+v_2(\xi)R'^2+v_4(\xi)R'^4+...,
\ee
where
\be
\xi\equiv \frac{r}{R(z)}
\ee
is the self-similarity variable.

These expressions are consistent with the jump conditions Equations~\eqref{bc1}~--~\eqref{bc4}. When Equations~\eqref{as1}~--~\eqref{as4} are plugged into the gas flow Equations~\eqref{gf1}~--~\eqref{gf3}, one gets a hierarchy of equations for successive orders. The calculation is straightforward, with only one subtlety. The operator $\partial_z$, when  applied to Equations~\eqref{as1}~--~\eqref{as4}, generates $R''$. The asymptotic expansion Equations~\eqref{as1}~--~\eqref{as4} will be consistent with the steady flow Equations~\eqref{gf1}~--~\eqref{gf3} iff one can also expand
\be
\frac{RR''}{R'^2}=b_0+b_1R'^2+b_2R'^4+...,
\ee
where $b$ are constants. 

Below we consider only the leading-order solution, but we must note that the first subleading order may also be interesting. At the first subleading order, a bow shock created by the rigid obstacle differs from a bow shock created by the stellar wind. And  bow shocks created by different stellar wind velocities differ from each other. 

At the leading order, all strong bow shocks have the same asymptotic. To leading order, 
\be
v=1.
\ee
This is because, to leading order, $v=1$ at the shock upstream and, to leading order, the Bernoulli equation gives $v={\rm const}$ along a streamline. Thus, asymptotically the bow shock downstream slows down to the ISM velocity. This doesn't mean that the bow shock becomes ``invisible'' at great distances from the head. Our bow shock is assumed infinitely strong, so the maximal density at any fixed arbitrarily large $z$ is still a factor of a few (4 for $\gamma =5/3$) greater than the unperturbed ISM density.
\begin{figure}[t]
    \centering
    \includegraphics[width=0.85\columnwidth]{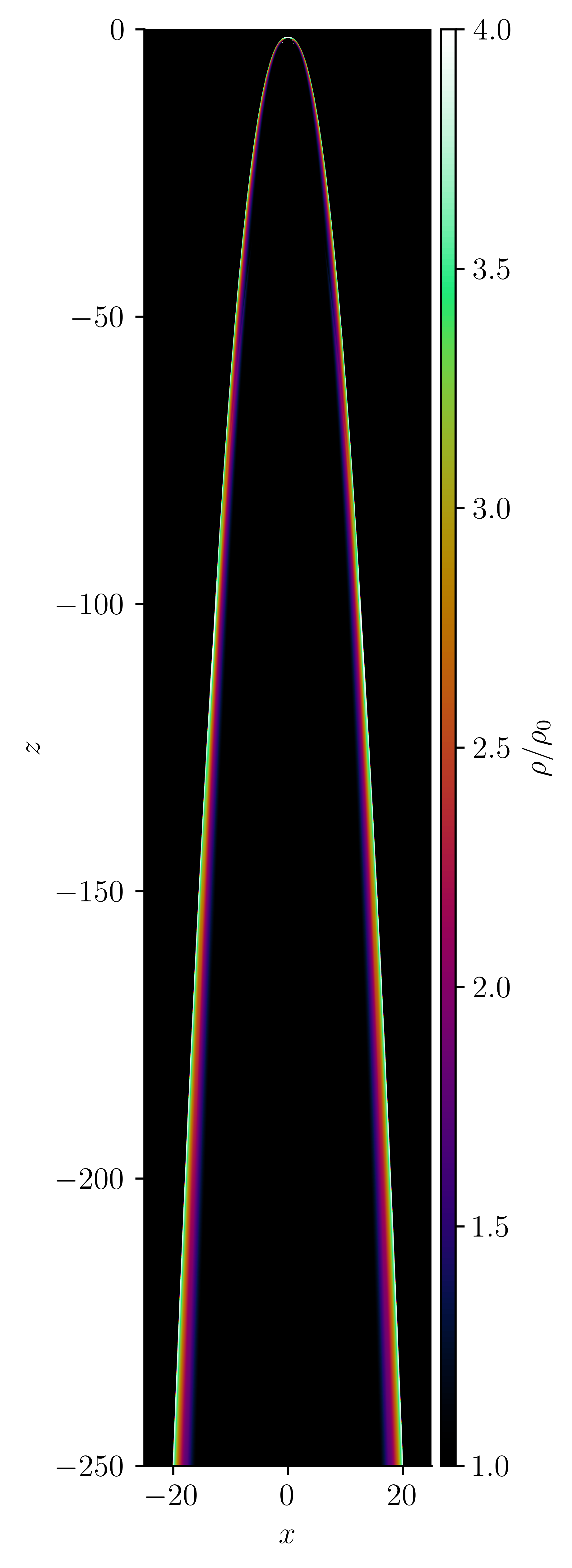}
    \caption{The full structure of the bow shock for a rigid obstacle plotted at $t = 1,000$. The bow shock wings extend to a maximum of 20  characteristic radii from the symmetry axis.}
    \label{fig:bow_rigid}
\end{figure}

With $v=1$, the system \eqref{gf1}~--~\eqref{gf3} reads
\begin{gather}
    \partial_z\rho+\frac{1}{r}\partial_r(r\rho u)=0,\label{gfr1}\\
    (\partial_z+u\partial_r)u=-\frac{1}{\gamma \rho}\partial_r(\rho c^2),\label{gfr2}\\
    (\partial_z+u\partial_r)\left(\ln c-\frac{\gamma -1}{2}\ln \rho\right)=0.\label{grf3}
\end{gather}
The boundary conditions, to leading order, are
\begin{gather}
    \rho=\frac{\gamma +1}{\gamma -1},\label{bcr1}\\
    c=\frac{\sqrt{2\gamma(\gamma -1)}}{\gamma +1}R',\label{bcr2}\\
    u=\frac{2}{\gamma +1}R'\label{bcr3}.
\end{gather}
Equations~\eqref{gfr1}~--~\eqref{bcr3} are isomorphic to a cylindrically symmetric blast wave; $z$ plays the role of time. 
An exact solution can immediately be deduced, analogous to the Sedov-von Neumann-Taylor solution \citep[][secion 106]{Landau+Lifshitz+1959}:
\begin{gather}
    R(z)\propto z^{1/2},\\
    \begin{align}
        \rho &= G(\xi) & u &= R^\prime U(\xi)\xi & c = R^\prime C(\xi)\xi
    \end{align}\\
    \xi^2=\frac{4}{\gamma^2-1}\left(\frac{\gamma -1}{\gamma (\gamma +1)}\right)^\frac{1}{\gamma}\frac{(U-\frac{1}{\gamma})^{(1-\frac{1}{\gamma})}}{U(\frac{2}{\gamma}-U)},\\
    \begin{split}
  G=\frac{\gamma+1}{\gamma-1}\left(\frac{\gamma (\gamma+1)}{\gamma-1}\right)^\frac{1}{\gamma}\left(\frac{\gamma (\gamma-1)}{2}\right)^\frac{2}{2-\gamma} \\
  \times \left(U-\frac{1}{\gamma}\right)^\frac{1}{\gamma}\left(\frac{\frac{2}{\gamma}-U}{1-U}\right)^\frac{2}{2-\gamma},
\end{split}\\
C^2 = \frac{\gamma -1}{2}\frac{U^2(1-U)}{U-\frac{1}{\gamma}}.
\end{gather}
We will see that these analytic expressions do accurately describe the far downstream of various bow shocks: created by the rigid obstacle, or by the stellar winds with different stellar wind velocity to the star velocity ratios. 

The focal length of the parabola, $f$,  is given by the energy deposition rate in the ISM frame\footnote{We thank Boaz Katz for pointing this out.}. For
\be\label{eq:rform}
R=2\sqrt{fz},
\ee
we have
\be\label{eq:norm} 
f^2=\frac{\beta^4}{32}\frac{\dot{M}_w(V^2+V_w^2)}{\rho_{\rm ISM} V^3},
\ee 
where $\dot{M}_w$ is the stellar wind mass loss rate, $V_w$ is the stellar wind velocity (relative to the star), $V$ is the velocity of the star, $\rho_{\rm ISM}$ is the ISM density; $\beta$ is the (2D) energy integral parameter (\cite{Landau+Lifshitz+1959}, section 106):
\be\label{eq:beta}
\beta^4\frac{\pi}{2}\int _0^1d\xi~\xi^3G\left(\frac{1}{2}U^2+\frac{1}{\gamma (\gamma -1)}C^2\right)=1.
\ee 
For $\gamma=7/5$, $\beta=1.01$, and for $\gamma=5/3$, $\beta =1.15$.

\begin{figure}[t]
    \centering
    \includegraphics[width=0.85\columnwidth]{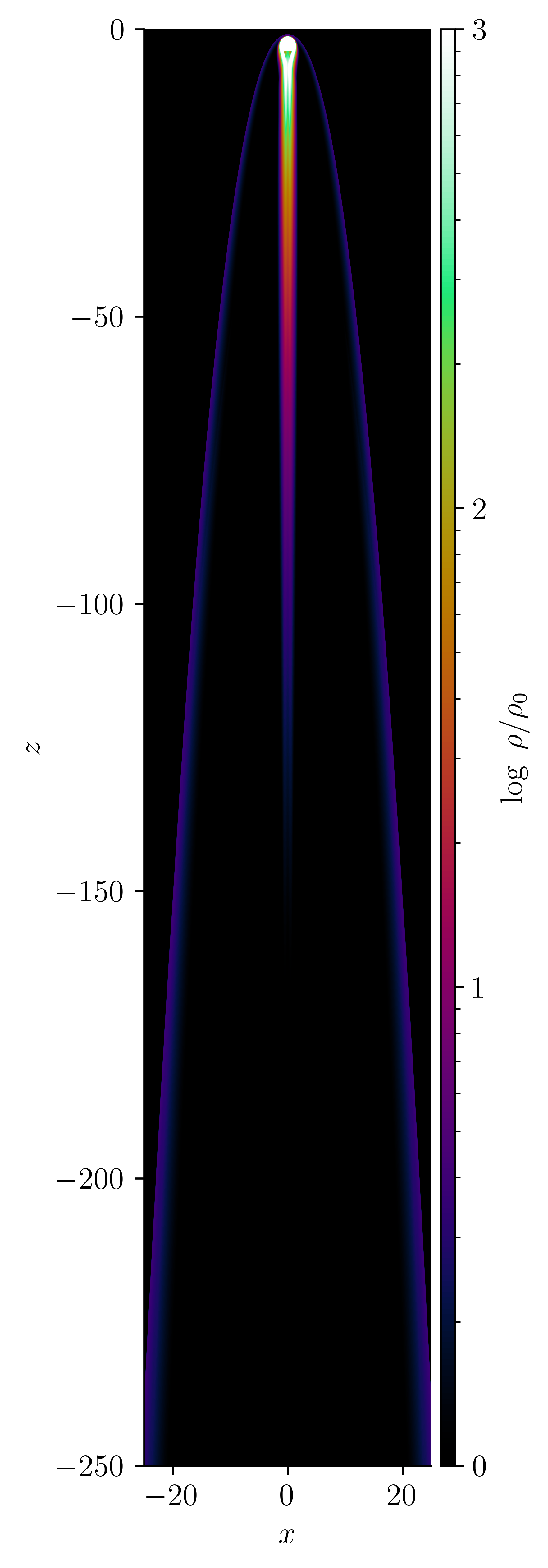}
    \caption{The full structure of the bow shock for an obstacle with $v_{\rm wind} = 0.1$ plotted at $t = 1,000$. The bow shock wings are wider than those of the rigid bow shown in Figure \ref{fig:bow_rigid} due to inner-outer wind interaction.}
    \label{fig:bow_v01}
\end{figure}
%
\section{Numerical Setup and Results}\label{sec:results}
\subsection{Numerical Setup}\label{sec:numerics}
The far asymptotic behavior of the bow shock can be solved numerically in 2D axisymmetry since turbulence near the head of the bow shock is washed out in the far downstream. However, turbulence is inherently a 3D phenomenon; therefore, we perform the numerical experiments in 2D to study the self-similar asymptotic discussed in Section \ref{sec:analytics}, and we use 3D to investigate the turbulence criteria described in Section \ref{sec:turb}.

We solve the Euler equations in conservative form: 
\begin{gather}
    \partial_t \rho + \nabla \cdot(\rho \textbf{v}) = 0,\label{eq:dcons}\\
    \partial_t (\rho\textbf{v}) + \nabla \cdot(\rho\textbf{v} \otimes \textbf{v} + p\mathbb{I}) = 0,\label{eq:mcons}\\
    \partial_t (\rho e_{\rm T}) + \nabla \cdot [(\rho e_T + p)\textbf{v}] = 0\label{eq:econs},
\end{gather}
 where $e_{\rm T} = p / [\rho(\gamma - 1)] + |\textbf{v}|^2/2$ is the total specific energy and $\mathbb{I}$ is the identity matrix. We solve Equations \eqref{eq:dcons} -- \eqref{eq:econs} using a second-order Godunov-type GPU-accelerated gas dynamics code entitled \texttt{SIMBI} \citep{DuPont+2023}. The initial condition is an ambient density of $\rho_\infty = 1$, ambient vertical velocity $v_{\infty} = -1$, and ambient Mach number $M_\infty = 10^4$ with $\gamma = 5/3$. The length scale is set by a characteristic obstacle size $\ell = 1$. For the rigid obstacle, $\ell$ is just the radius. For the windy obstacles, $\ell$ is the stand-off distance of the outermost shock. All calculations are done in the rest frame of the obstacle.

We first compute the solution for the solid obstacle. To achieve this, we invoke the immersed boundary method as described in \cite{peskin_2002} where we place a spherical and impermeable boundary surface on top of the grid. We treat the rigid body as a discrete set of markers on the grid with discrete forcing as similarly described in \cite{Gronskis+2016}.  This method perfectly captures a rigid body to avoid erroneous ablation effects and Rayleigh-Taylor fingers that would penetrate and distort the obstacle shape. For the wind, we continuously inject mass into a volume such that the stand-off density is $\rho_{\rm s} > \rho_\infty (v_\infty / v_{\rm wind})^2$, where $v_{\rm wind}$ is the ballistic wind velocity. Let $ \scriptr = \sqrt{r^2 + (z - z_0)^2}$ where $z_0$ is the obstacle's vertical offset, then the wind is 
\begin{equation}\label{eq:mdot}
    \rho_{\rm wind}(\scriptr \leq \ell) = \rho_{\rm s}(\scriptr / \ell) ^ {-2}.
\end{equation}
As a parameter study, we investigate the set of wind speeds $v_{\rm wind} = \{0.1, 1, 10 \}$ and their impacts on the far downstream flow.

In 2D, we use an axisymmetric cylindrical mesh with square zones. The vertical domain spans $z\in [0, -250]$ and the radial domain spans $r\in[0, 25]$. We ensure at least 102 zones across the obstacle's cross section for which the solution is converged, which implies a resolution of $12,800$ vertical zones by 1,280 radial zones. We choose $z_0 = -2.5$. We run all simulations until a time $t = 1,000$ to ensure statistical saturation. Simulations use the generalized minmod slope limiter with $\theta_{\rm PLM}$, a numerical diffusivity factor, set to 1.5 for near-minimal diffusion and more robust preservation of contact surfaces.

As a check for theoretical robustness, we revisit this calculation 
in full 3D. A limitation of 3D is that we cannot feasibly resolve a Cartesian grid of $12,800 \times 2560 \times 2560$ zones, so we instead simulate the windy obstacles with $v_{\rm wind} = \{0.1, 1\}$ on a coarse-grained 3D grid. We cannot feasibly check for numerical convergence of the asymptotic bow shock shape in 3D, but we will use this as a litmus test to: (a) see whether there are differences in the saturated flow when going from 2D to 3D; and (b) investigate the assertion made in Section \ref{sec:turb} that the tangential discontinuity is turbulent near the head of the obstacle. For (a) we ensure 20 zones across the obstacle cross section on a grid that spans $z\in[0,-40]$ and $x,y \in [-20,20]$. For (b) we ensure 200 zones across the obstacle cross section on a grid that spans $z\in[0,-4]$ and $x,y \in [-2,2]$.

\subsection{Results}
\begin{figure}[t]
    \centering
    \includegraphics[width=\columnwidth]{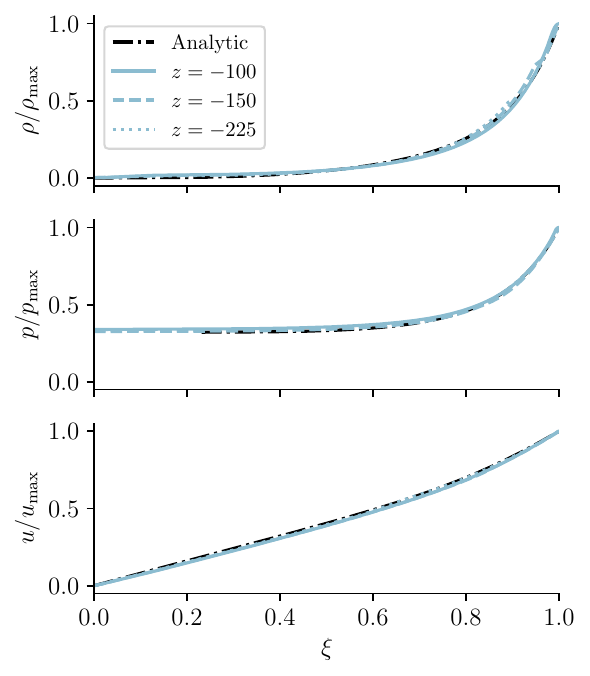}
    \caption{Shown are radial slices of $\rho, p,$ and $u$ at time $t = 1,000$ plotted as functions of the self-similarity variable $\xi$ for the rigid obstacle. $\xi = 1$ marks the location of the strong bow shock. The dash-dotted black curve marks the theoretical prediction. The blue solid, dashed, and dotted blue curves mark the simulation results at $z = -100, -150,$ and $-225$. }
    \label{fig:rigid_xi}
\end{figure}
In Figures \ref{fig:bow_rigid} and \ref{fig:bow_v01} we plot the two edge cases considered in this work. Namely, the full structure of the bow shock for the rigid body and the windiest body, i.e, $v_{\rm wind} = 0.1$, respectively. In Figure \ref{fig:bow_rigid} the density field behind the rigid body is evacuated (i.e., $\rho \ll 1$), a testament to the robustness of the immersed boundary method invoked to model the boundary layer. In Figure \ref{fig:bow_v01}, the density field only gradually decreases to values $\rho \lesssim 0.1$ at large distances $|z| \geq 150$. This hints at the fact that the windy obstacles approach the asymptotic solutions slower than the perfectly rigid obstacle since the downstream is polluted by turbulent streams of ablated material. 

Moreover, the self-similar theory and numerical simulation show near perfect agreement for the rigid body as shown by Figure \ref{fig:rigid_xi}. There, we plot the quantities $\rho$, $p$, and $u$ as functions of the self-similar variable, $\xi$. We consider $100\ell$ as the nominal threshold for the far downstream and plot each hydrodynamic quantity in logarithmic intervals further downstream to better showcase the self-similar nature of the bow shock extent.
\begin{figure}[t]
    \centering
    \includegraphics[width=\columnwidth]{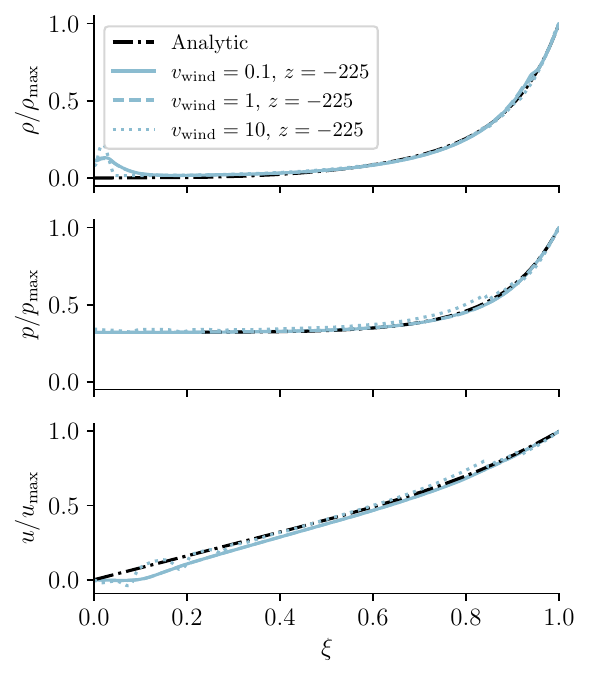}
    \caption{Shown are radial slices of $\rho, p,$ and $u$ at time $t = 1,000$ plotted as functions of the self-similarity variable $\xi$ for the windy obstacles. $\xi = 1$ marks the location of the strong bow shock. The dash-dotted black curve marks the theoretical prediction. The blue solid, dashed, and dotted curves mark the simulation results for obstacles with wind speeds 0.1, 1, and 10, respectively. All values are plotted at $z = -225$. }
    \label{fig:wind_xi}
\end{figure}

Since in reality, the more realistic body flying through the ISM is likely a star with some wind, we plot in Figure \ref{fig:wind_xi}  $\rho$, $p$, and $u$ for the wind speeds $v_{\rm wind} = \{0.1, 1, 10\}$ at a fixed distance of $225\ell$. Although some turbulence is present in the far downstream as evidenced by the density bumps near the axis, the overall structure of the bow shock shape still matches nicely with the exact asymptotic derived in the previous section. 

\begin{figure*}[t!]
    \gridline{\fig{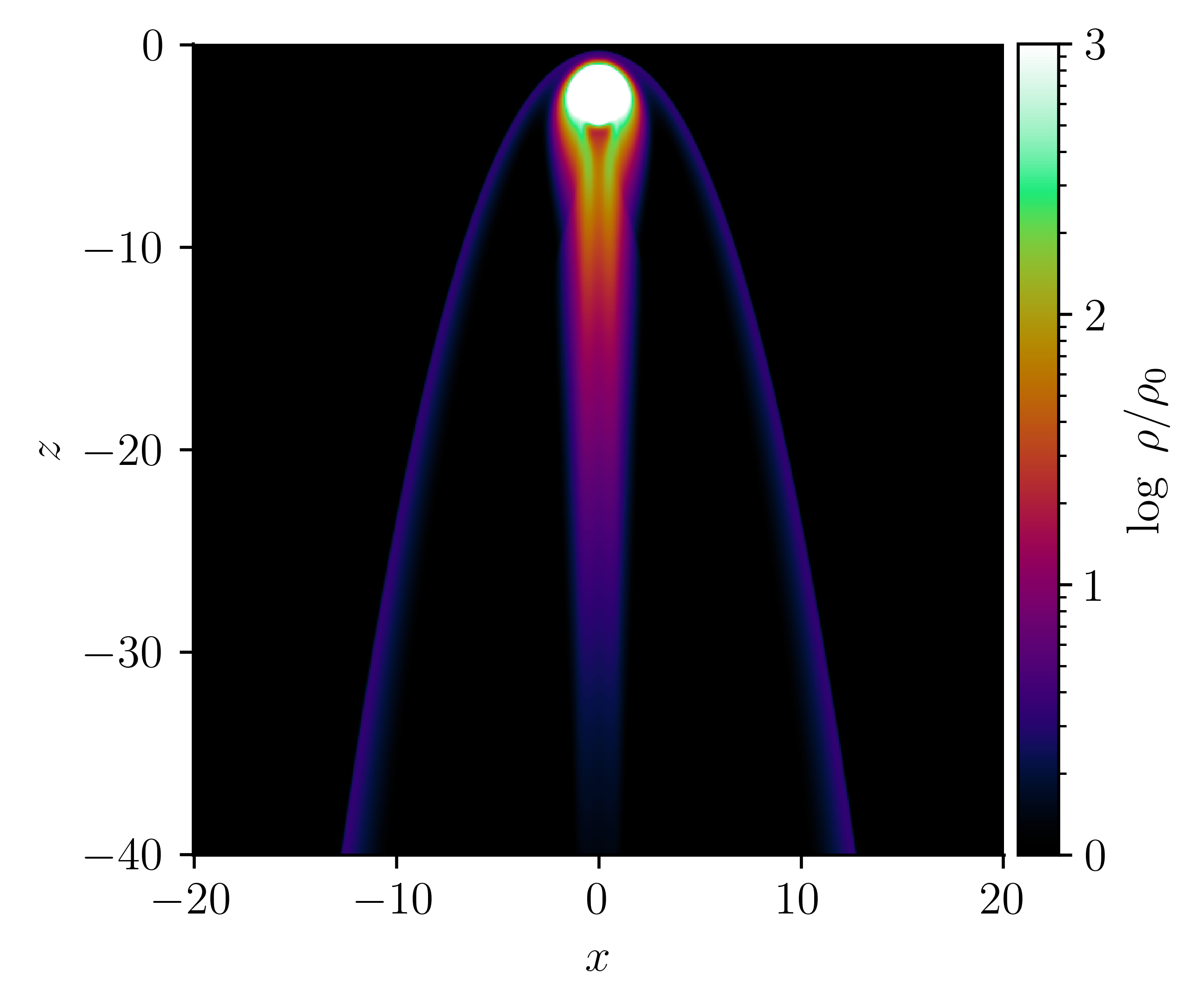}{0.5\textwidth}{(a)}\fig{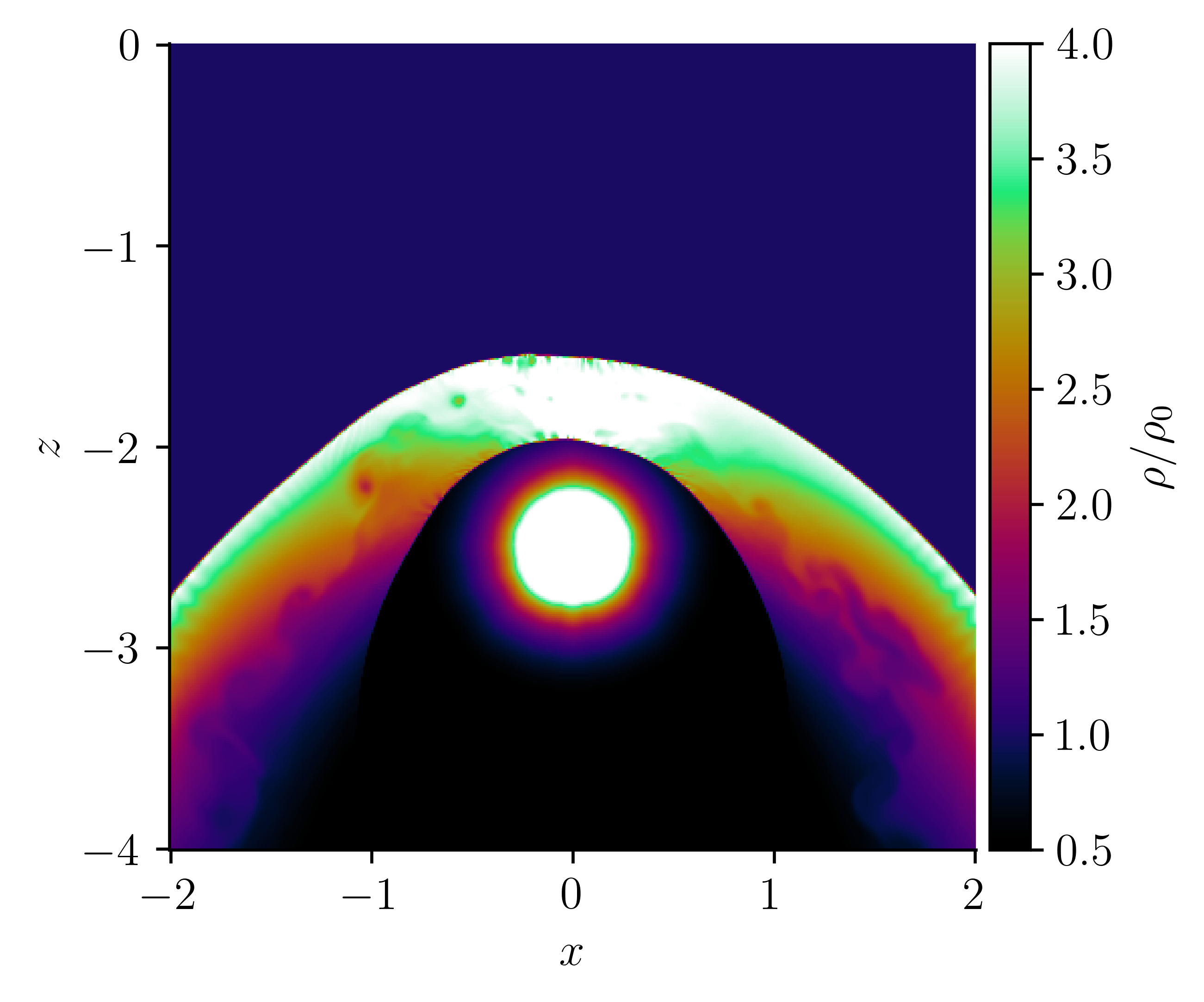}{0.5\textwidth}{(b)}}
    \caption{The bow shock with infinite Mach number when done in full 3D. We plot $v_{\rm wind} = 0.1$ case in panel (a) at $t = 1,000$ and $v_{\rm wind} = 1$ in panel (b). In panel (a) we plot the logarithm of the density field, which shows similar signs of decay along the axis as depicted in the higher resolution 2D runs. 
    In panel (b) we plot the density field for an obstacle with $v_{\rm wind} = 1$.
    The forward shock is relatively stable, but behind it lies the unstable tangential discontinuity between the inner and outer winds. The instability manifest as turbulence as predicted in Section \ref{sec:turb}.
    }
    \label{fig:3d_wind}
\end{figure*}

The 3D simulations shown in Figure \ref{fig:3d_wind} are of the windy obstacles with $v_{\rm wind} = 0.1$ in panel (a) and $v_{\rm wind} = 1$ in panel (b). Though the 3D simulations were performed with a coarser resolution than the 2D axisymmetric case, panel (a) still shows the same characteristics of the high resolution 2D cases, i.e., the gradual dissipation of the wind material along the axis of the obstacle. Furthermore, panel (b) is a zoomed in region very near the obstacle with $v_{\rm wind} = 1$, and we can see the forward shock is relatively stable while axisymmetry is broken by the turbulent contact discontinuity just as we discussed in Section \ref{sec:turb}. Thus, our theory of non-magnetic astrophysical bow shocks is supported by a numerical experiment.  
\section{Discussion}\label{sec:discussion}
D
This work has shown that the far asymptotic structure of the bow shock with infinite Mach number is self-similar and and can be described analytically. After derivation of the exact asymptotic solution, we confirm the theory with high-resolution hydrodynamical simulations for a rigid obstacle and a windy obstacle. The rigid obstacle shows near perfect agreement with the theory with a maximal percent difference of 7\%. The windy obstacle shows rough agreement with the theory, but this is to be expected since the presence of an inner wind introduces turbulence. As the obstacle gets decelerated, the bow shock shape will transition from a parabola to the canonical Mach cone with opening angle $\theta_M = \sin^{-1}(1/M)$. Using Equation \eqref{eq:rform} in conjunction with the Mach angle, we arrive at
\begin{eqnarray}\label{eq:ztrans}
        z_{\rm Mach} &=& 4f(M^2 - 1),\\
        R_{\rm Mach} &=& 4f(M^2 - 1)^{1/2},
\end{eqnarray}
as the Mach cone transition coordinates. This implies that the assumption that the bow shock remains so strong far from the obstacle is valid in range $R \ll z \ll 4fM^2$ for high Mach numbers.  

Although inspired by the work of \YS{}, we have independently developed an exact, analytic solution to the bow shock with infinite Mach number without making any assumptions about the resultant solid of revolution a priori. Due to its exactness and more direct experimental validation, we suggest that our self-similar bow shock theory is more robust methodologically than what is described in \YS{}.

Having constructed and verified this theory expounding the asymptotic shape of non-magnetic astrophysical bow shocks, the astrophysical implications remain unknown. That is to say, the physical systems for which our model applies remains to be validated and we speculate on the model's applicability below. For example, an interesting astrophysical application of our model is the study of pulsar wind kinematics. For many decades, observers have modeled the H$\alpha$ bow shocks around pulsars using the \W{} solution despite the fact that these H$\alpha$ bow shocks are non-radiative \citep{Kulkarni+1988,vanKerkwijk+2001, Brownsberger+2014,Romani+2022}. Since our model is both adiabatic and asymptotically analytic up to first order, we suggest that our solutions coupled with the focal length in Equation \eqref{eq:norm} might provide more stringent constraints on the pulsar wind energetics when modeling far from the obstacle head. An important followup would be to revisit this with realistic pulsar-ISM wind speed ratios in the simulations to better model the bow shock shape nearer to the obstacle head.

Moreover, this model may also be applicable to runaway supernova (SN) ejecta blobs that outrun the SN shock front and develop bow shocks like what has been observed in the filamentary structures of the Cassiopeia A SN remnant \citep{Fesen+2006} or the bidirectional ejecta blobs that might develop  from explosions of axisymmetric Roche-lobe distorted Type Ib/c SN progenitors \citep{DuPont+2024}.

Another potential use case for our model is constraining the sources behind quasi-periodic eruptions \citep[QPEs;][]{Miniutti+2019,Giustini+2020,Arcodia+2021,Chakraborty+2021}. Some models suggest that QPEs are stars that plunge into active galactic nuclei \citep[AGN; e.g.,][]{Lu+Quataert+2023,Linail+Metzger+2023,Tagawa+Haiman+23}. The supersonic star creates a bow shock whose thermal energy is radiation dominated and this model is used to explain the radiative properties of some QPEs. In the future, we plan to utilize our theory to explore the fundamental problem of stars plunging through AGN disks in order to glean some of the physics of the stellar entry and exit. This should provide constraints on the \'a la mode QPE models proposed.\\

    \noindent We thank the referee for providing useful comments which led to the improvement of this work. The authors thank Tamar Faran,  Zoltan Haiman, Boaz Katz, and Yuri Levin for useful discussions. MD acknowledges a James Author Fellowship from NYU's Center for Cosmology and Particle Physics, and thanks the LSST-DA Data Science Fellowship Program, which is funded by LSST-DA, the Brinson Foundation, and the Moore Foundation; his participation in the program has benefited this work.


\bibliography{refs}{}
\bibliographystyle{aasjournal}



\end{document}